\documentclass[conference]{IEEEtran}
\IEEEoverridecommandlockouts
\usepackage{cite}
\usepackage{amsmath,amssymb,amsfonts}
\usepackage{graphicx}
\usepackage{textcomp}
\usepackage{xcolor}
\usepackage{booktabs}
\usepackage{multirow}
\usepackage{tabularx}
\usepackage{url}

\def\BibTeX{{\rm B\kern-.05em{\sc i\kern-.025em b}\kern-.08em
    T\kern-.1667em\lower.7ex\hbox{E}\kern-.125emX}}
\begin{document}

\title{Institutionalizing Trust in Data Exchange: Trust-Anchor Model of the Data Credit Guarantee Organization and Its Simulation-Based Comparison}

\author{\IEEEauthorblockN{Hiroshi Mano}
\IEEEauthorblockA{
\textit{EverySense Japan Inc.}
\\Tokyo, Japan
\\h.mano@every-sense.com} 
\and
\IEEEauthorblockN{Chie Sugiyama}
\IEEEauthorblockA{
\textit{Data Society Alliance}
\\Tokyo, Japan
\\c.sugiyama@data-society-alliance.org} 
\and
\IEEEauthorblockN{Teruaki Hayashi}
\IEEEauthorblockA{
\textit{Hitotsubashi University}
\\Tokyo, Japan
\\teruaki.hayashi@r.hit-u.ac.jp} }

\maketitle

\begin{abstract}
Data trading markets that span industries and regions are expanding, but the implicit trust that supports data use within an organization does not carry over to cross-domain exchange. This paper examines the Data Credit Guarantee Organization (DCGO), a proposal to institutionalize trust through third-party verification and certification, and asks what it means to trust data, what such certification can support, and under what conditions DCGO outperforms existing trading models. We decompose trust in data into four layers: authenticity of entity, authenticity and integrity of data, rights and conditions of use, and history and auditability. A third party can certify authenticity but not the correctness of content, and DCGO is positioned as a trust anchor spanning existing marketplaces and data spaces. We then extend the Law Merchant model with three data-specific properties, namely imperfect ex-post detection, cheap pseudonyms, and non-rivalry of harm, and compare bilateral trading with DCGO by agent-based simulation. DCGO keeps the buyer's payoff near its theoretical upper bound regardless of the market environment, whereas bilateral trading loses even when fraud is always detected. Identity verification proves to be a prerequisite for record sharing, and pre-verification is the only function that limits the spread of harm. On the seller side, honest sellers join voluntarily whenever the price premium exceeds the fee. Lock-in and a critical mass for diffusion, however, call for fee governance and support for entry.
\end{abstract}


\section{Introduction}
\label{sec:intro}
\subsection{Data Trading Markets and the Fragmentation of Trust}
\label{sec:intro-fragmentation}

With the development of AI and big data, data have come to be regarded as a strategic asset that determines the performance of AI, decision making, and service delivery. Global private investment in generative AI reached USD 33.9 billion in 2024~\cite{stanford2025}, and the market for data marketplaces is projected to grow from about USD 1.5 billion in 2024 to about USD 5.7 billion by 2030~\cite{gvr2024}. This trend is pushing up the demand for data whose quality, rights clearance, and provenance have been verified, and data trading markets for discovering, acquiring, and reusing data across industries and regions are expanding. They take two broad forms: data marketplaces, in which providers list ``data products'' in a catalog and users search, contract, and use them in the cloud, and data spaces, federated ecosystems in which participants keep data under their own control and share them on the basis of common identifiers, connection methods, usage policies, and governance rules.

What these markets have in common is that they combine third-party trust functions, such as identity verification of organizations, electronic signatures and certificates issued by trusted third parties (TTPs), access control, contracts and terms of use, and audit logs, with the operational governance of each operator. A signature or certificate, however, shows only that a specific entity issued specific information or that alteration can be detected. It does not guarantee that the data represent real-world facts, that the provider holds sufficient rights, or that the intended use is permitted. Which evidence is verified, how certificates are revoked and corrected, and who bears responsibility in a dispute are determined by the governance of each marketplace or data space operator.

This operator-specific governance allows flexible design according to the business characteristics of each market, but it fragments the evidence of trust when it must be carried between markets. A certification, evaluation, or audit result obtained in one market is not necessarily recognized in another, so identity verification, rights verification, quality verification, and audit preparation are repeated for the same provider or the same data at every connection point. This raises the cost of participation for providers and the time and uncertainty from discovery to use for users. The growth of data trading markets therefore requires, beyond more capable individual markets, evidence of trust that can be used across markets and a common mechanism for verification and the demarcation of responsibility.

\subsection{Positioning of DCGO and Research Questions}
\label{sec:intro-dcgo}
The Data Society Alliance (DSA) in Japan has advocated a mechanism that secures trust through third-party verification and certification in order to realize Data Free Flow with Trust (DFFT)~\cite{dfft}. In fiscal year 2026, the construction of a highly trusted data collaboration infrastructure centered on the Data Credit Guarantee Organization (DCGO) was adopted under the Cabinet Office's BRIDGE program~\cite{bridge2025}, and efforts toward its social implementation are under way.

DCGO is not intended to replace existing data marketplaces or data spaces. Rather, as a trust anchor that spans them, it provides common evidence formats, verification procedures, assurance levels, and rules for revocation, correction, and audit, so that evidence of trust can be carried between markets and data spaces in a verifiable form while preserving the expertise and flexibility of each operator. DCGO thus does not substitute for the internal governance of individual markets; by resolving the cross-market information asymmetry and fragmentation of trust that are hard to solve within a single market, it complements existing governance and may accelerate the growth of data trading through the expansion of data supply, demand, reuse, and the scope of trading.

The focus of this paper therefore shifts from whether trust exists in existing markets to a question of institutional design: which trust functions should be shared across markets so as to complement existing governance while lowering the barriers to participation and the information asymmetry in data trading? We address it through three research questions.
\begin{itemize}
\item \textbf{RQ1:} What does it mean to trust data?
\item \textbf{RQ2:} What, and to what extent, can a mechanism of verification and certification such as DCGO support?
\item \textbf{RQ3:} Under what conditions does DCGO work effectively compared with existing trading models?
\end{itemize}

Our contributions are threefold. First, we decompose trust in data into four layers, namely the authenticity of entity, the authenticity and integrity of data, rights and conditions of use, and history and auditability, and define what trustworthy data are (RQ1). Second, we map the architecture of DCGO onto the four layers and clarify the scope and boundary of what third-party certification can support (RQ2). Third, we model a data market with information asymmetry as an agent-based simulation and compare bilateral trading with DCGO, showing the conditions under which DCGO benefits both buyers and sellers and which layer is responsible for which effect (RQ3).


\begin{figure*}[t]
\centering
\includegraphics[width=\textwidth]{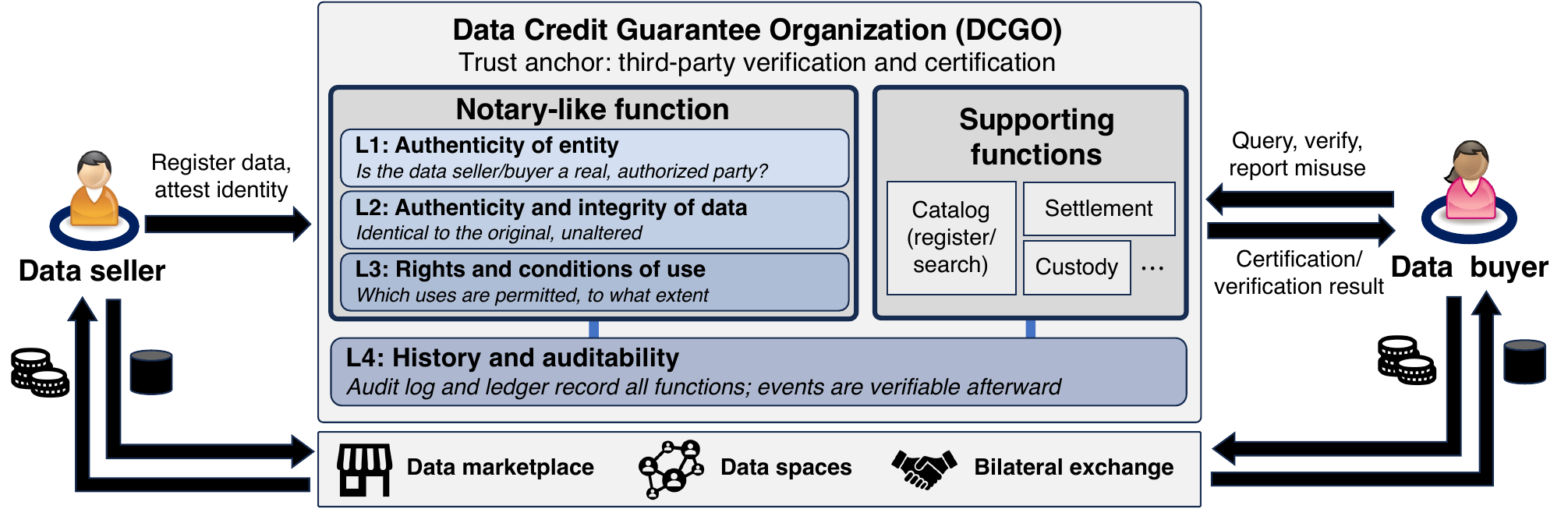}
\caption{Architecture of the Data Credit Guarantee Organization (DCGO). DCGO acts as a trust anchor between data sellers and buyers: a notary-like function certifies the authenticity of entity (L1), the authenticity and integrity of data (L2), and rights and conditions of use (L3), and the audit log and ledger (L4) record all functions. Data and payments flow through existing marketplaces, data spaces, or bilateral exchange, not through DCGO.}
\label{fig:architecture}
\end{figure*}

\section{Related Initiatives and Research}
\label{sec:related}
\subsection{Data Marketplaces and Data Spaces}
\label{sec:rw-markets}

Existing data trading markets fall into three design axes. The first is the market platform type, which catalogs data products and brokers search, purchase, delivery, and settlement. AWS Data Exchange~\cite{aws} and Snowflake Marketplace~\cite{snowflake} are commercial examples in which users subscribe to data products from within the provider's cloud environment, and EverySense Pro in Japan offers catalog search, negotiation, delivery, and settlement for inter-company trading while the operator confines itself to brokerage~\cite{everysense}. The second is the data space type, in which participants retain control over their data and use them across organizations on the basis of common connection methods, access rights, and usage rules. Catena-X, an industry-led data space of the automotive industry, exchanges data directly between participants through connectors based on the IDS reference architecture, with conditions of use negotiated and enforced as policies between connectors~\cite{catenax_a,catenax_b}. The third is the rights certificate type, which expresses the scope of permitted actions on data as rights and trades those rights. In the data usage rights trading market that DSA is testing, the conditions of use defined by the provider (access, method of use, redistribution, and period of use) are embodied as a ``data usage rights certificate'' bound to the data and traded~\cite{dsa_rights}. Real services combine these axes; the Kyoto Data Marketplace, a catalog site for public and private data~\cite{kyoto}, can connect a catalog with a data space and incorporate usage rights and contract terms into it.

\subsection{Standardization of Data Trading Systems and Data Spaces}
\label{sec:rw-standards}

For data spaces, the International Data Spaces Association (IDSA) maintains the IDS reference architecture and is standardizing the Dataspace Protocol for communication between connectors~\cite{idsa_dsp}, Gaia-X provides a trust framework for evaluating the trustworthiness of participants and services~\cite{gaiax}, and industrial data spaces such as Catena-X build on these. In Japan, the Digital Architecture Design Center of the Information-technology Promotion Agency (IPA) is developing ODS-RAM (Open Data Spaces Reference Architecture Model)~\cite{ods}, and DSA promotes DATA-EX, an initiative that provides domain-independent common technologies and standards for building data spaces, in alignment with the ODS specifications~\cite{dsa_ods}.

IEEE 3800-2024, ``IEEE Standard for a Data-Trading System: Overview, Terminology, and Reference Model,'' defines the terminology, overview, and reference model of a data trading system and describes the relationships among data providers, data users, trading systems, and data space operators with common concepts, thereby providing a reference frame for interoperability between different trading systems~\cite{ieee3800}. It does not, however, guarantee that individual data are correct as real-world facts, that the provider holds all rights, or that no violation of conditions of use occurs after the trade. We therefore position DCGO on top of this reference frame as a trust anchor that provides, as a third party, evidence of trust, verification procedures, and assurance levels shared across trading systems and data spaces built on it.

\subsection{Provenance, Description, and Authenticity of Data}
\label{sec:rw-tech}

Individual technologies that support trust in data have also accumulated. W3C PROV provides a standard model for describing the generation, transformation, and attribution of data~\cite{prov}, documentation frameworks such as Datasheets for Datasets make explicit the purpose of collection, composition, and constraints of machine learning data~\cite{gebru2021}, and, in addition to electronic signatures, timestamps, and hash chains, watermarking techniques verify after the fact the origin and unauthorized use of data and trained models~\cite{agrawal2002,uchida2017}. These technologies mainly support the authenticity and integrity of data and history and auditability among the four layers in Section~\ref{sec:framework}, but none of them answers the institutional questions of who verifies, who guarantees, and who bears responsibility upon revocation or dispute. This paper takes them as given and positions DCGO as a mechanism that bundles them and provides assurance across markets.

\subsection{Economic Analysis and Simulation Studies}
\label{sec:rw-econ}

Information asymmetry in data markets is a market-for-lemons problem~\cite{akerlof1970}: if buyers cannot tell quality before purchase, high-quality sellers leave and overall quality declines. Classical institutional remedies are the Law Merchant model, in which a third party that keeps records of cheating sustains honest trade~\cite{milgrom1990}, and the certification intermediary that inspects and discloses quality~\cite{lizzeri1999}. Online, however, cheap pseudonyms undermine record-based institutions~\cite{friedman2001}, and data are a credence good whose quality is unknown until use~\cite{darby1973} and whose easy copying lets a single fraud harm many users at once.

Beyond work on data pricing and mechanism design~\cite{pei2022,agarwal2019}, agent-based studies of data markets are advancing. Yamamoto and Hayashi compared reputation systems in a manufacturing data market with agents whose utilities were estimated by Q-learning and inverse reinforcement learning, showing that the reputation design affects the alignment of price and quality~\cite{yamamoto2025}. Sashihara et al.\ proposed a multi-agent framework with large language models as buyer and seller agents~\cite{sashihara2025} and used it to analyze how institutional designs such as transaction prohibitions affect the growth and decline of data markets~\cite{sashihara2026}. None of these studies, however, incorporates third-party verification and certification as a market function.

In summary, standardization provides interoperability, provenance and authenticity technologies provide individual pieces of evidence, and economic analysis and simulation provide means to understand market dynamics, but no study has cut across these fields to decompose what it means to trust data or to ask, from the viewpoint of institutional design, under what conditions a third-party verification and certification mechanism works compared with existing trading models. This paper fills that gap by decomposing trust into four layers, mapping DCGO onto them, and testing the conditions under which DCGO works by an agent-based simulation that extends the Law Merchant model with data-specific properties.

\section{Trust Framework and DCGO Architecture}
\label{sec:framework}

\subsection{Four Layers Supporting Trust in Data}
\label{sec:layers}

Unlike a paper certificate or a physical product, data are intangible: copying them does not consume the original, and identical copies can be made in large numbers in a short time. Whether the data in front of us are ``genuine'' is therefore hard to judge by looking at them. A file name that claims to be the original does not guarantee that it is the first record, and even if the figures are transcribed accurately, judgment may be wrong if the measuring instrument was not calibrated. Moreover, even when authenticity can be verified, if the data were redistributed beyond the scope of the license, the user runs the risk of unexpected liability. Trust in data thus has several faces, and a single certificate cannot cover all of them.

In this paper we define trust in data as a state in which the following four questions, each concerning a different object of verification, can be answered.

\noindent\textbf{Authenticity of entity (L1):} Do the provider and the user exist, and are they entities with legitimate authority to provide and use the data?

\noindent\textbf{Authenticity and integrity of data (L2):} Are the delivered data identical to the presented original and free from tampering?

\noindent\textbf{Rights and conditions of use (L3):} Which of the actions such as viewing, copying, processing, AI training, redistribution, overseas provision, storage, and deletion are permitted, and to what extent?

\noindent\textbf{History and auditability (L4):} Are events such as acquisition, registration, viewing, processing, transfer, redistribution, and deletion recorded so that they can be traced and verified afterward?

The four layers verify different objects, namely the parties (L1), the data themselves (L2), the permitted actions (L3), and the events over time (L4), and cannot substitute for one another. Data that have not been tampered with cannot be used with confidence if they are sold by an unauthorized entity (L2 holds but L1 does not). Data received from a legitimate provider and verified as untampered may still not be usable for AI training if the license does not permit it (L1 and L2 hold but L3 does not), or cannot be relied upon for quality judgment if the method of acquisition and the processing history are unknown (L1 and L2 hold but L4 does not).

A point to emphasize is the existence of a boundary of what can be certified. The certification of L2 by hash values, electronic signatures, e-seals, and timestamps gives evidence that a given piece of data was issued by a specific entity and has not been altered since. It does not, however, guarantee that the content of the data agrees with real-world facts. Evaluating the correctness of content requires combining evidence that belongs to the other layers, such as the managing entity (L1), the declared purpose and scope of use (L3), and calibration records, installation conditions, and measurement methods (L4), and judging it against the intended use.

What the four-layer framework aims at is therefore not a ``guarantee of correctness'' but assembling the evidence needed for judgment and making previously invisible risks assessable. Trustworthy data are not data that contain no errors; they are data whose origin and limitations can be verified and which can be reasonably evaluated against their intended use. This is our answer to RQ1.

\subsection{DCGO as a Trust Anchor}
\label{sec:dcgo-anchor}

Of the four layers, L1 and L2 have been realized individually by existing technologies such as corporate identifiers and electronic signatures, and L3 and L4 have been handled partially by contracts and operations. As described in Section~\ref{sec:intro-fragmentation}, however, these are established separately by each operator, and in collaboration across domains and borders it is difficult for the trading parties alone to establish them every time: providers fear use beyond the intended purpose and leakage, and users doubt authenticity, quality, and rights. Under this two-sided information asymmetry, the higher the quality of data, the less likely they are to reach the market, and the market turns into a market for lemons.

DCGO is a proposal in which a third party independent of the trading parties verifies and certifies the four layers against common criteria. We call it a \emph{trust anchor} in the sense that evidence verified once by DCGO is recognized, in the same form and at the same assurance level, across the markets and data spaces that refer to DCGO; the certification of the four layers is thereby bundled into ``portable trust'' that accompanies a transaction wherever it takes place. Although the name ``credit guarantee'' may evoke monetary compensation for losses, the core of the proposal is to neutrally verify the evidence and rules required for data trading and thereby raise the trustworthiness of transactions.

Figure~\ref{fig:architecture} shows the architecture at the present stage. DCGO is placed between data providers and users, and at its core is a notary-like function that certifies (i) the authenticity of entity (L1), (ii) the authenticity and integrity of data (L2), and (iii) the unity of the data and the contractual conditions of their use (L3). Combined with this core are supporting functions: (iv) a data catalog (registration and search), (v) custody, (vi) settlement, and (vii) support by AI agents for discovery, comparison of conditions, negotiation, and receipt. The audit log and ledger (L4) record the operations of all these functions, so that any certification can be traced back to the evidence on which it was based. Data and payments themselves flow through existing marketplaces, data spaces, or bilateral exchange, not through DCGO. ``Notary'' here does not refer to the notary public system under current law but to a notary-like function that technically and neutrally certifies authenticity and related properties.

Three distinctions are important for understanding DCGO. First, DCGO is not a data repository. Rather than aggregating all data centrally, a federated design is assumed in which each participant keeps its data and only certification information and metadata are linked; for highly confidential data, conditions can be confirmed without disclosing the content. Second, DCGO is not a trading site. Before a market function that attaches a price and sells, a notary-like function that verifies who can provide what under which conditions is needed. Without this foundation of trust, high-quality data will not gather even with market functions in place. Third, DCGO does not replace existing data infrastructures. It is a layer that superimposes a cross-cutting trust verification function on the collaborations established in each industry and use case and on the mechanisms that connect them.

In summary, what DCGO can support is the verification and certification of evidence for L1 to L4 and its portability across markets; guaranteeing the correctness of content, storing data, and brokering sales are outside its scope, as is judging whether the data fit a particular purpose, which requires the user's context. This is our answer to RQ2. Section~\ref{sec:results} examines the validity of this positioning by showing which layers play roles that cannot be substituted by the others (RQ3).

\section{Models of DCGO for Simulation}
\label{sec:model}

\subsection{Objective}
\label{sec:objective}

To answer RQ3, that is, under what conditions DCGO works effectively compared with existing trading models, we ask whether using DCGO is beneficial in the long run for both buyers and sellers even after paying its fees, and we identify the parameter region in which this holds. ``Beneficial'' has two meanings. For buyers, it means that the net payoff exceeds that under the existing trading model on average over a sufficiently long period. For sellers, it means that an honest seller obtains more profit by joining DCGO than by not joining, so that joining is chosen voluntarily. We verify the former with a buyer-side simulation and the latter with a seller-side simulation, and define the region in which both hold as the \emph{condition under which DCGO works}. By removing the functions of DCGO one at a time, we also identify which of the four layers is responsible for which effect.

We use agent-based simulation rather than equilibrium analysis because the market considered here has dynamic properties that are difficult to treat analytically: participants turn over, buyers learn from finite experience, and a single fraud harms many buyers at once. The simulation is not calibrated with real data; what it provides is a qualitative condition, namely which parameters matter and in which direction, rather than a quantitative prediction.

\subsection{The Law Merchant Model and Data-Specific Extensions}
\label{sec:lawmerchant}

We model the market after the Law Merchant model~\cite{milgrom1990}, which explains why honest trade was sustained at medieval fairs where trading partners changed frequently. A third party without enforcement power keeps records of past cheating and answers queries before a trade. If the cycle of \emph{query, record, and exclude} is individually rational, honest trade is sustained even in a large market of one-shot encounters where bilateral reputation does not work. The ledger of DCGO plays this role in data trading.

To this model, we add three properties specific to data as a good. First, fraudulent data cannot be recognized until they are used: tampering, forged provenance, and defective rights clearance surface only later and only partially, as malfunctions of analysis results or trained models. We represent this as \emph{imperfect ex-post detection}, in which fraud is discovered after the trade only with probability~$\delta$. The smaller $\delta$, the fewer frauds are recorded and the weaker record-based exclusion becomes. This corresponds to a credence good~\cite{darby1973}.

Second, an excluded fraudulent seller can return with a new account. Unlike a medieval merchant, whose identity is fixed, a new ID is cheap to obtain in digital trading. We represent this as the social cost of \emph{cheap pseudonyms}~\cite{friedman2001}: an excluded fraudulent seller returns as a new ID with probability~$\rho$. This property separates trading models without identity verification from DCGO, which blocks re-entry.

The third is the \emph{non-rivalry of harm}. Because data are easy to copy, a single undetected fraudulent seller can harm many buyers in the same period. We place no upper limit on the quantity a seller can supply, so that multiple buyers can choose the same seller at once, and measure this by \emph{victims per fraud}. This property raises the social cost of delayed detection and hence the value of verification before the trade.

Finally, DCGO verifies the authenticity of data before the trade, a function absent from the Law Merchant model that corresponds to a \emph{certification intermediary}~\cite{lizzeri1999}, a third party that inspects sellers' quality and discloses it to buyers. In our model, pre-verification detects and excludes fraudulent data before the trade with probability~$p_{\mathrm{det}}$. DCGO is thus the composition of two classical third-party mechanisms, the law merchant (records and queries) and the certification intermediary (inspection before the trade), with querying and recording automated through its catalog and ledger rather than left to the voluntary action of merchants. The ablation study separates what each mechanism contributes and what happens when either is missing.

\subsection{Trading Models}
\label{sec:institutions}

We represent each trading model by the presence or absence of three functions of DCGO: sharing of fraud records among buyers (L4), identity verification that binds records to a seller and blocks re-entry under a new ID (L1), and pre-verification of data before the trade (L2). The trading model with none of the three is called \emph{Bilateral}, in which there is no third party and a buyer judges its counterpart only from its own experience, and the one with all three is called \emph{DCGO}. For the ablation study we also set \emph{DCGO$-$ID}, which removes only identity verification, and \emph{DCGO$-$Check}, which removes only pre-verification (Table~\ref{tab:institutions}). Because the four trading models differ only in these functions within the same model, the effect of each function can be isolated.

Rights and conditions of use (L3) is not handled, because L1, L2, and L4 all matter for whether a buyer can trust a seller at the time of the trade, whereas L3 concerns the buyer's behavior after the trade, such as use beyond the intended purpose, and requires a different model that includes post-purchase behavior, which we leave as a future extension.

Who bears the usage fee is a design variable: a fee borne by buyers raises their participation threshold directly, whereas a fee borne by sellers may reduce the incentive of honest sellers to join. Buyers bear the fee in the buyer-side simulation, and sellers bear it in the seller-side simulation, where buyers pay only a small fee $c_b$ for certified data.

\begin{table}[t]
\centering
\caption{Trading models compared and the functions each one has}
\label{tab:institutions}
\footnotesize
\setlength{\tabcolsep}{4pt}
\begin{tabular}{@{}lccc@{}}
\toprule
Trading model & \shortstack{Sharing of fraud\\records (L4)} &
\shortstack{Identity\\verification (L1)} &
\shortstack{Pre-verification\\(L2)} \\
\midrule
Bilateral   & $\times$   & $\times$   & $\times$   \\
DCGO        & \checkmark & \checkmark & \checkmark \\
DCGO$-$ID    & \checkmark & $\times$   & \checkmark \\
DCGO$-$Check & \checkmark & \checkmark & $\times$   \\
\bottomrule
\end{tabular}
\end{table}

\subsection{Agent Model}
\label{sec:agents}

The market has $N$ sellers and $M$ buyers. Seller $j$ has a fixed type $\theta_j \in \{0,1\}$: a \emph{fraudulent seller} ($\theta_j=1$) always sells fraudulent data and an \emph{honest seller} ($\theta_j=0$) always sells correct data, and the share of fraudulent sellers is $\phi$. A buyer obtains value $v$ from good data and suffers loss $L$ from fraudulent data. The price $p$ is fixed, and $L$, $p$, and the DCGO fee $c$ are expressed as ratios to $v=1$.

In every round, each seller leaves the market with probability $\mu_s$ and each buyer with probability $\mu_b$, and they are replaced by the same number of new entrants: a new seller is fraudulent with probability $\phi$, a new buyer starts with no trading experience, and the records concerning a departed seller (blacklist entries and detection) leave with it. This turnover represents the Law Merchant situation in which the chance of meeting the same partner again is small.

A trading model is represented by a tuple $(s,\iota,\pi,c)$: three binary \emph{switches} corresponding to the three functions in Section~\ref{sec:institutions}, namely sharing of fraud records $s$ (L4), identity verification $\iota$ (L1), and pre-verification $\pi$ (L2), each equal to $1$ if the trading model has the function and $0$ otherwise, and the usage fee $c \ge 0$ that a buyer pays for each trade. If a buyer never receives fraudulent data, its payoff from one trade is $v-p-c$, which we call the \emph{theoretical upper bound} of the buyer's net payoff ($0.60$ for Bilateral with $c=0$ and $0.55$ for DCGO with $c=0.05$).

\subsection{A Simulation Round}
\label{sec:round}

Each round $t$ consists of the following seven steps.

\noindent\textbf{(Step 1) Turnover.} Sellers and buyers are replaced with the probabilities described above.

\noindent\textbf{(Step 2) Participation decision.} Each buyer $i$ holds a \emph{subjective fraud rate} $\tilde r_i^{\,t}$, its estimate of the share of fraudulent data in the market, formed from the frequency of fraud detected in its own past trades and, when records are shared, in those of other buyers. The buyer participates if the expected payoff from a single trade is positive:
\begin{equation}
(1-\tilde r)\,v - \tilde r L - p - c > 0
\;\Longleftrightarrow\;
\tilde r < \tau \equiv \frac{v-p-c}{v+L},
\label{eq:threshold}
\end{equation}
where $\tau$ is the threshold fraud rate above which a buyer stops participating; the higher the fee $c$, the lower the threshold. A non-participating buyer re-enters on a trial basis with a small probability $\varepsilon$, so that a buyer that once left can notice that the market has improved.

\noindent\textbf{(Step 3) Seller choice.} A participating buyer chooses one seller uniformly at random among the sellers not on the blacklist $B_{ij}^{t}$ it can refer to. There is no upper limit on the quantity a seller can supply, so several buyers may choose the same seller in the same round and, if it is fraudulent, all of them suffer the loss (non-rivalry of harm).

\noindent\textbf{(Step 4) Pre-verification (DCGO only).} If the chosen seller is fraudulent, the fraud is detected before the trade with probability $p_{\mathrm{det}}$ and the seller is immediately blacklisted. Because verification is performed regardless of whether the trade is completed, the buyer pays the fee $c$ even when fraud is detected.

\noindent\textbf{(Step 5) Payoff.} A buyer that completes a trade receives $v-p-c$ for good data and $-L-p-c$ for fraudulent data; a buyer blocked by pre-verification receives $-c$.

\noindent\textbf{(Step 6) Ex-post detection and memory update.} If the trade is completed and the seller is fraudulent, the fraud is discovered with probability $\delta$; the seller is blacklisted and the buyer's subjective fraud rate is updated. The switch $s$ determines whether these records are shared among all buyers or remain private to the buyer that detected it. With the shared blacklist and fraud rate written as $b_j^{t}$ and $r^{t}$, and the private ones of buyer $i$ as $b_{ij}^{t}$ and $r_i^{t}$, buyer $i$ refers to
\begin{equation}
B_{ij}^{t} = s\,b_j^{t} + (1-s)\,b_{ij}^{t},
\qquad
\tilde r_i^{\,t} = s\,r^{t} + (1-s)\,r_i^{t}.
\label{eq:sharing}
\end{equation}
With $s=0$, a buyer can avoid only the sellers by which it was itself harmed.

\noindent\textbf{(Step 7) Pseudonym reset.} A blacklisted fraudulent seller re-enters the market as a new ID with probability $\rho$ in trading models without identity verification ($\iota=0$); with identity verification ($\iota=1$), it is permanently excluded.

\subsection{Metrics}
\label{sec:metrics}

The \emph{buyer's net payoff} is the payoff per buyer per round, that is, the value of good data minus the loss from fraudulent data, the price, and the usage fee. The \emph{participation rate} is the share of buyers that participated in that round. The \emph{contamination rate} is the share of fraudulent data among the data purchased in that round. \emph{Victims per fraud} is the number of buyers harmed per seller that sold fraudulent data in that round, which captures the spread of harm due to the copyability of data. In the seller-side simulation we additionally observe the share of honest sellers that are certified (the \emph{adoption rate}) and the profit per honest seller per round.

\subsection{Buyer-Side and Seller-Side Simulations (Sim~1 and Sim~2)}
\label{sec:sim12}

To make the results easier to interpret, we run two simulations in which one side is fixed. In \emph{Sim~1}, the buyer-side simulation, the types and behavior of sellers are fixed, only buyers learn and decide following Section~\ref{sec:round}, the trading model is applied uniformly to the whole market, and the usage fee is paid by buyers.

In \emph{Sim~2}, the seller-side simulation, buyers are fixed and a voluntary DCGO certification is installed on top of a Bilateral base market, in which a buyer avoids only the sellers whose fraud it has itself noticed, records are not shared, pseudonym resets are not prevented, and buyers are replaced at rate $\mu_b$ with empty blacklists. A seller that joins DCGO undergoes identity verification (and cannot re-join once revoked) and pre-verification at every sale, pays a fee $c_s$ per unit sold, and in exchange sells certified data at a price premium $\eta$. Buyers choose among certified sellers whenever they exist and pay $c_b$ for certified data. An honest seller with production cost $k$ earns $p-k$ per unit if uncertified and $p+\eta-c_s-k$ if certified. Sellers do not optimize the decision to join; each seller compares its average profit over the last 10 rounds with that of a randomly chosen seller and imitates the membership status of the more profitable one. A fraudulent seller may also imitate joining, but it is stopped by pre-verification at its first sale and loses its certification. Newly entering sellers start uncertified.

\subsection{Experimental Settings}
\label{sec:settings}
Table~\ref{tab:params} lists the definitions and default values of the parameters used in the simulations. One run consists of 300 rounds, and 30 runs are performed with different random seeds. Reported values are averages over the last 100 rounds, excluding the initial transient, across the 30 runs.

\begin{table*}[t]
\centering
\caption{Parameters and their default values}
\label{tab:params}
\small
\setlength{\tabcolsep}{6pt}
\renewcommand{\arraystretch}{1.4}
\begin{tabularx}{\textwidth}{@{}lXl@{}}
\toprule
Symbol & Meaning & Default \\
\midrule
$N$, $M$ & Number of sellers, number of buyers & 100, 200 \\ \hline
$\phi$ & Share of fraudulent sellers among sellers & 0.3 \\ \hline
$\delta$ & Probability that a buyer notices fraud after buying fraudulent data & 0.3 \\ \hline
$\rho$ & Probability that a blacklisted fraudulent seller re-enters via a pseudonym reset & 0.2 \\ \hline
$\mu_s$, $\mu_b$ & Share of sellers / buyers replaced by new participants per round & 0.02, 0.02 \\ \hline
$v$, $L$, $p$ & Value of good data, loss from fraudulent data, price ($v=1$ as the unit) & 1, 2, 0.4 \\ \hline
$p_{\mathrm{det}}$ & Probability that pre-verification blocks fraudulent data & 0.9 \\ \hline
$c$ & Usage fee of the trading model, paid by the buyer per trade (Sim~1 only) & Bilateral: 0, DCGO: 0.05 \\ \hline
$\varepsilon$ & Probability that a non-participating buyer participates on a trial basis (Sim~1 only) & 0.05 \\ \hline
$k$ & Cost for an honest seller to produce one unit of good data; 0 for fraudulent sellers (Sim~2 only) & 0.2 \\ \hline
$\eta$ & Price premium of certified data (Sim~2) & 0.10 \\ \hline
$c_s$, $c_b$ & DCGO fees: $c_s$ is paid by the seller, $c_b$ by the buyer per unit of certified data (Sim~2 only) & 0.05, 0.05 \\
\bottomrule
\end{tabularx}
\end{table*}


\begin{figure*}[t]
\centering
\includegraphics[width=\textwidth]{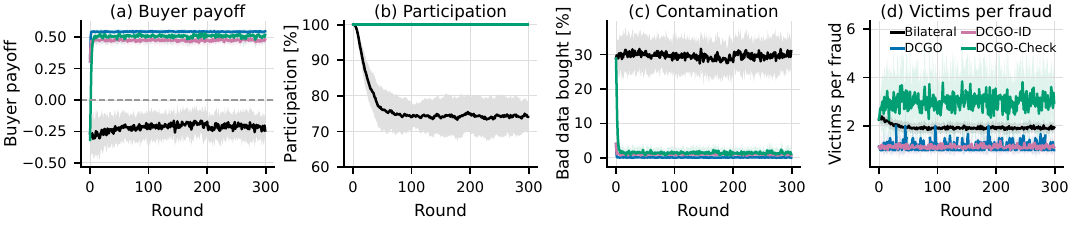}
\caption{Time series by trading model: (a) buyer's net payoff,
(b) participation rate, (c) contamination rate, and (d) victims per fraud. Solid lines are the means over 30 seeds; bands are mean $\pm$ standard deviation.}
\label{fig:sim1-ts}
\end{figure*}

\begin{figure}[t]
\centering
\includegraphics[width=7cm]{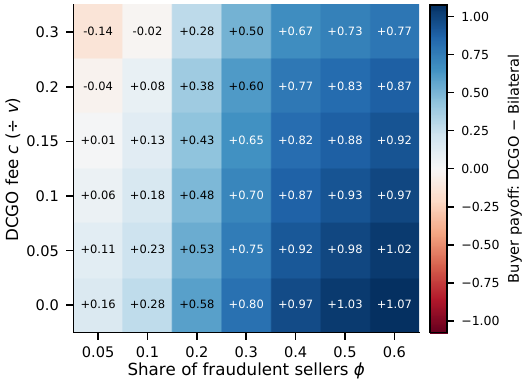}
\caption{Difference in the buyer's net payoff. Darker blue means DCGO is more favorable; darker red means Bilateral is more favorable.}
\label{fig:sim1-phase}
\end{figure}

\begin{figure*}[t]
\centering
\includegraphics[width=\textwidth]{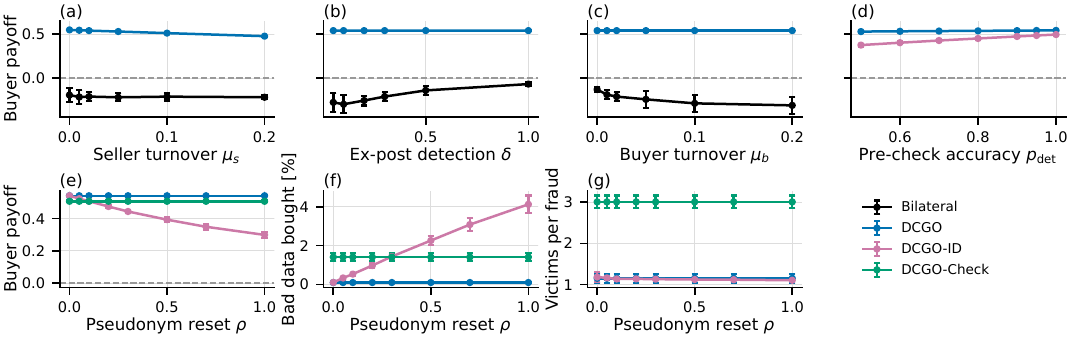}
\caption{Sensitivity analysis of Sim~1. Buyer's net payoff under Bilateral and DCGO when (a) the seller turnover $\mu_s$, (b) the ex-post detection probability $\delta$, and (c) the buyer turnover $\mu_b$ are varied one at a time from their default values, and (d) under DCGO and DCGO$-$ID when the pre-verification accuracy $p_{\mathrm{det}}$ is varied. (e) Buyer's net payoff, (f) contamination rate, and (g) victims per fraud as functions of the pseudonym-reset probability $\rho$.}
\label{fig:sim1-sens}
\end{figure*}

\begin{figure*}[t]
\centering
\includegraphics[width=17cm]{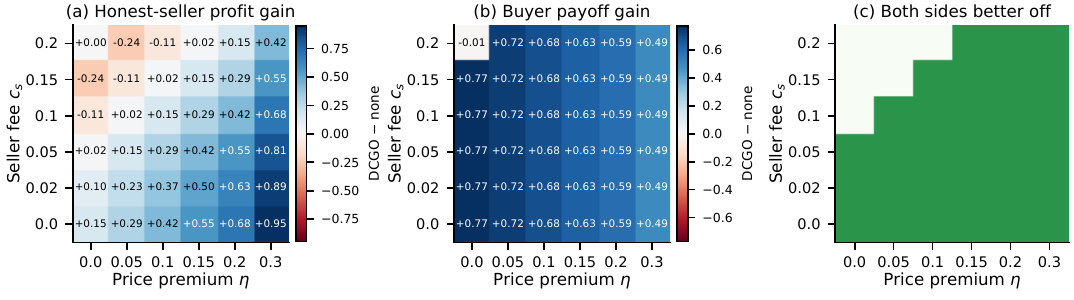}
\caption{For each combination of the price premium $\eta$ (horizontal
axis) and the seller-side fee $c_s$ (vertical axis): (a) gain in the
profit of honest sellers, (b) gain in the buyer's net payoff, and (c)
combinations in which both are positive. Gains are differences from the
market without DCGO.}
\label{fig:sim2-phase}
\end{figure*}

\begin{figure*}[t]
\centering
\includegraphics[width=15cm]{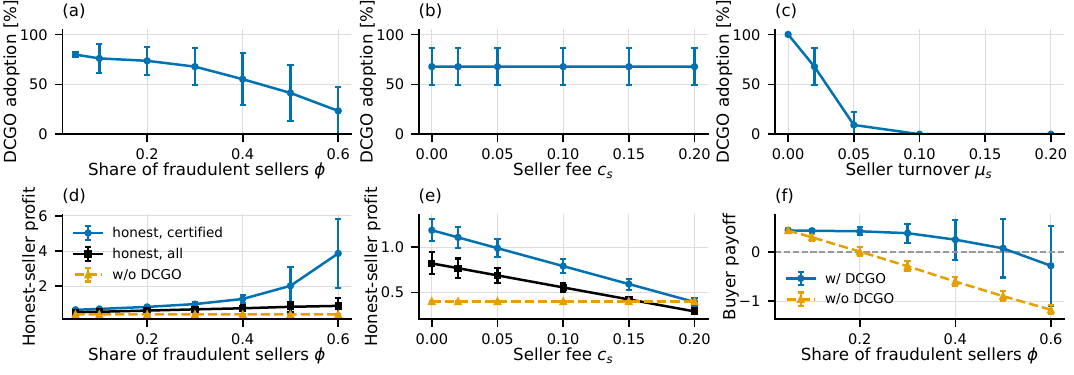}
\caption{Sensitivity analysis of Sim~2: DCGO adoption rate of honest sellers when (a) the share of fraudulent sellers $\phi$, (b) the seller-side fee $c_s$, and (c) the seller turnover $\mu_s$ are varied. Profit of honest sellers as a function of (d) $\phi$ and (e) $c_s$, and (f) buyer's net payoff as a function of $\phi$.}
\label{fig:sim2-sens}
\end{figure*}

\section{Results and Discussion}
\label{sec:results}

\subsection{Buyer-Side Simulation (Sim~1)}
\label{sec:results-sim1}

Under Bilateral, buyers lose from the early rounds on (Table~\ref{tab:sim1}, Fig.~\ref{fig:sim1-ts}(a)). The contamination rate stays at about 30\%, nearly equal to the share of fraudulent sellers $\phi=0.3$ (Fig.~\ref{fig:sim1-ts}(c)): because of seller turnover and pseudonym resets, the market as a whole cannot accumulate the knowledge of which sellers are fraudulent, even though each buyer builds its own blacklist. Under DCGO, in contrast, the buyer's payoff almost reaches the theoretical upper bound of $0.55$. Even after paying the usage fee, it is $0.75$ higher than under Bilateral, and the contamination rate is kept down to $0.1\%$. Notably, the participation rate under Bilateral stops falling at 74\% (Fig.~\ref{fig:sim1-ts}(b)): buyers notice only 30\% of fraudulent data after the fact and cannot realize that they are losing. In a market for data in which fraud is hard to notice, a high participation rate is therefore not evidence that the market is sound; the market must be assessed by both participation and payoff.

Figure~\ref{fig:sim1-phase} shows the difference in the buyer's net payoff between DCGO and Bilateral for each combination of $\phi$ and the usage fee $c$. DCGO is inferior only when there are almost no fraudulent sellers ($\phi \le 0.1$) \emph{and} the fee is high ($c \ge 0.2$). In almost the entire remaining region DCGO yields the higher payoff, by $+0.50$ at $\phi=0.3$ even when $c=0.3$. The difference increases almost linearly in $\phi$ and decreases almost linearly in $c$. In the sensitivity analysis (Fig.~\ref{fig:sim1-sens}), the DCGO payoff stays within $0.53$--$0.55$ whichever of $\phi$, the seller turnover $\mu_s$, the ex-post detection probability $\delta$, $\rho$, and the pre-verification accuracy $p_{\mathrm{det}}$ is varied; the only parameter that moves it is the fee $c$. The Bilateral payoff, in contrast, turns negative at around $\phi \approx 0.2$ and remains negative even when fraud is always noticed (Fig.~\ref{fig:sim1-sens}(b)) or pseudonym resets are absent, and it falls as the buyer turnover $\mu_b$ rises (Fig.~\ref{fig:sim1-sens}(c)), because new buyers do not know who is fraudulent.

In summary, the Bilateral payoff declines as the market environment deteriorates, whereas the DCGO payoff is constant regardless of the environment, with its upper bound lowered only by the fee. The conditions under which DCGO works are thus governed by two quantities, the strength of the incentive to cheat $\phi$ and the cost of certification $c$. This is the buyer-side answer to RQ3.

The ablation study (Fig.~\ref{fig:sim1-sens}(e)--(g)) compares DCGO, DCGO$-$ID, and DCGO$-$Check as $\rho$ is varied. At $\rho=0$, when the identity of sellers is completely fixed, DCGO$-$ID achieves the same outcome as DCGO: this is the world of the Law Merchant model, in which fixed identity was an implicit premise and record sharing alone sustains honest trade. As $\rho$ increases, however, the payoff of DCGO$-$ID decreases monotonically, from $0.48$ at $\rho=0.2$ to $0.39$ at $\rho=0.5$ and $0.30$ at $\rho=1$, and the contamination rate rises from $0.1\%$ to $4.1\%$: under cheap pseudonyms~\cite{friedman2001}, the exclusion of fraudulent sellers remains incomplete without identity verification, even with record sharing and pre-verification. Under DCGO and DCGO$-$Check, which have identity verification, none of the payoff, the contamination rate, and the number of victims changes with $\rho$. Identity verification is the function that effectively sets $\rho$ to $0$. Without it the history-and-auditability layer cannot keep functioning. Among the four layers, L1 is therefore a prerequisite for L4.

A further result concerns the number of victims per fraud (Table~\ref{tab:sim1}, Fig.~\ref{fig:sim1-ts}(d)), which is larger in trading models that have fewer undetected fraudulent sellers: $1.9$ under Bilateral versus $3.0$ under DCGO$-$Check. The more record sharing shrinks the set of fraudulent sellers not yet blacklisted, the more buyers concentrate on the few that remain, and because data can be copied without limit, one fraudulent seller sells to all of them in the same round. This is the non-rivalry of harm added to our model, captured by the victims-per-fraud metric. Only DCGO and DCGO$-$ID, which have pre-verification, keep the number of victims at around one by stopping fraud before the trade. The pre-verification accuracy $p_{\mathrm{det}}$ has almost no effect on the buyer's payoff (Fig.~\ref{fig:sim1-sens}(d)) but does affect the number of victims.

Therefore, reducing the number of fraudulent sellers and reducing the harm caused by a single fraud are distinct goals: the former is achieved by record sharing and identity verification, and the latter only by pre-verification. This is why L2 is necessary among the four layers.

\newcolumntype{Y}{>{\centering\arraybackslash}X}
\begin{table}[t]
\centering
\caption{Comparison of trading models with the default parameters \\($\pm$ is the standard
deviation across seeds)}
\label{tab:sim1}
\footnotesize
\begin{tabularx}{\columnwidth}{@{}lYYY@{}}
\toprule
Trading model & Buyer's net payoff & Contamination rate & Victims per fraud \\
\midrule
Bilateral     & $-0.21 \pm 0.05$ & 29.7\% & 1.9 \\
DCGO          & $+0.54 \pm 0.00$ & 0.1\%  & 1.2 \\
DCGO$-$ID     & $+0.48 \pm 0.01$ & 1.0\%  & 1.1 \\
DCGO$-$Check  & $+0.51 \pm 0.01$ & 1.4\%  & 3.0 \\
\bottomrule
\end{tabularx}
\end{table}

\subsection{Seller-Side Simulation (Sim~2)}
\label{sec:results-sim2}

Table~\ref{tab:sim2} shows that the profit of honest sellers rises from $0.40$ to $0.69$ (about $1.7$ times), and to $0.99$ (about $2.5$ times) for certified sellers alone. Adoption spreads through imitation and settles at 68\% of honest sellers. Fraudulent sellers that join are stopped by pre-verification at their first sale and lose their certification, so the contamination rate falls from $29.5\%$ to $2.1\%$, and the buyer's net payoff turns from $-0.29$ to $+0.39$. The sellers' gain is not carved out of the buyers' share: the loss caused by fraudulent data disappears from the market as a whole, and that amount is divided between the two sides through the premium $\eta$.

As the share of fraudulent sellers $\phi$ increases (Fig.~\ref{fig:sim2-sens}(a)(d)(f)), the profit of certified honest sellers reaches $2.0$ at $\phi=0.5$, five times the level without DCGO, whereas the adoption rate declines and stays at 23\% at $\phi=0.6$: the more fraudulent sellers there are, the greater the value of joining, but the fewer the opportunities to meet a profitable member to imitate. This dilemma, that the more valuable the market, the harder it is for DCGO to spread naturally, rebounds on buyers: at $\phi=0.6$ a contamination rate of 26\% remains and the buyer's payoff is $-0.28$ even with DCGO (Fig.~\ref{fig:sim2-sens}(f)), a large improvement over $-1.18$ without DCGO but still a loss. Buyers benefit only once adoption on the seller side has spread sufficiently.

The condition under which both sides benefit is simple: the premium $\eta$ roughly exceeds the fee $c_s$ (Fig.~\ref{fig:sim2-phase}(c)). The sellers' gain is almost proportional to $\eta-c_s$ (Fig.~\ref{fig:sim2-phase}(a)), and the buyers' gain, although it decreases with $\eta$, is positive wherever DCGO has spread and still $+0.49$ at $\eta=0.3$ (Fig.~\ref{fig:sim2-phase}(b)). The boundary of the region is therefore set by the seller-side condition: as long as the fee can be passed on to the price, buyers accept a premium of up to about $\eta=0.3$, which is funded by the payoff difference of about $0.75$ measured in Sim~1.

There are, however, two side effects and a diffusion condition. First, because buyers prefer certified sellers, uncertified honest sellers (mainly new entrants) sell almost nothing (Table~\ref{tab:sim2}), so the gap between members and non-members becomes extremely large. Second, once DCGO has spread, the adoption rate stays at 68\% even when $c_s$ is raised to $0.2$ (Fig.~\ref{fig:sim2-sens}(b)), although the profit of certified sellers ($0.39$) then falls below the level without DCGO ($0.40$) (Fig.~\ref{fig:sim2-sens}(e)); sellers are locked in because uncertified sellers cannot sell, which implies that the fee must be restrained by governance rather than by market competition. Third, diffusion has a critical mass: even under the default setting, in 2 of the 30 seeds the initial members (5\%) were lost through turnover before imitation could multiply them, and with fast seller turnover ($\mu_s \ge 0.05$) this becomes the norm (Fig.~\ref{fig:sim2-sens}(c)), so that prior registration of large providers or support at entry is likely to be necessary in markets with intense entry. This is the seller-side answer to RQ3; together with the buyer-side answer, it gives the conditions under which DCGO works.

\begin{table}[t]
\centering
\caption{Markets without DCGO (Bilateral) versus voluntary
DCGO}
\label{tab:sim2}
\footnotesize
\setlength{\tabcolsep}{3pt}
\begin{tabular}{@{}lcccc@{}}
\toprule
DCGO &
\shortstack{Adoption\\rate} &
\shortstack{Honest-seller profit\\(all / certified / uncertified)} &
\shortstack{Contamination\\rate} & \shortstack{Buyer's\\net payoff} \\
\midrule
w/o & n/a & 0.40 / n/a / 0.40 & 29.5\% & $-0.29$ \\
w/ & 68\% & 0.69 / 0.99 / 0.03 & 2.1\%  & $+0.39$ \\
\bottomrule
\end{tabular}
\end{table}

\section{Conclusion}
\label{sec:conclusion}

\subsection{Summary}
\label{sec:summary}

This paper took up the Data Credit Guarantee Organization (DCGO), which performs third-party verification and certification, as a mechanism that makes trust, previously closed within each operator, portable across markets, and addressed three research questions. For RQ1, trust in data is a state in which four mutually non-substitutable layers can be answered: the authenticity of entity (L1), the authenticity and integrity of data (L2), rights and conditions of use (L3), and history and auditability (L4); trustworthy data are not error-free data but data whose origin and limitations can be verified against their intended use. For RQ2, DCGO is a trust anchor in which a third party verifies and certifies the evidence of the four layers and makes it portable across markets and data spaces; guaranteeing the correctness of content, storing data, and brokering sales are outside its scope. For RQ3, the DCGO payoff on the buyer side stays close to the theoretical upper bound regardless of the market environment and falls below bilateral trading only when fraudulent sellers are almost absent and the fee is high; on the seller side, honest sellers join voluntarily and both sides gain whenever the price premium exceeds the fee, with lock-in and a critical mass for diffusion as side effects.

The conditions under which DCGO works are thus described by two axes, the strength of the incentive to cheat and the cost of certification, and among the four layers the authenticity of entity and pre-verification play roles that record sharing alone cannot substitute. The priority in introducing DCGO is therefore not to upgrade market functions but to establish a foundation of trust that includes identity verification and pre-verification, and to keep its cost, through governance, at a level that providers and users can accept.

\subsection{Limitations and Future Work}
\label{sec:future}

We kept the data price fixed, because endogenous pricing would mix price formation with the formation and collapse of trust and make it harder to read which function of which trading model is at work. Our model also limits fraud to a single kind, namely data that cause a loss to the buyer; distinguishing fraud in authenticity, which DCGO targets, from errors in content, and modeling the propagation of contamination to secondary data, remain future work. In addition, we fixed either buyers or sellers; moving both sides at the same time and examining buyers' voluntary choice of certification are left for the future.

Among the four layers, rights and conditions of use (L3) was not handled. Violations of L3 appear as the buyer's behavior after the trade, such as use beyond the intended purpose or redistribution, so buyers must be modeled as agents that keep acting after purchase, and the effect of usage rights certificates and of the detection and revocation of violations must be evaluated. The support by AI agents for discovery, comparison of conditions, and negotiation (Section~\ref{sec:dcgo-anchor}) corresponds to the part of the trading procedure that we fixed; incorporating it would show how DCGO certification is referred to in agreements between agents. Modeling L3 and introducing AI agents are the next steps toward covering all four layers.

Our simulation also targets a single market and does not model the cross-market portability of trust itself; a model with multiple markets in which DCGO certifications are or are not mutually recognized is future work. Finally, the parameters are not calibrated with real data, so our findings are limited to qualitative conditions. The next step is to obtain the fraud rate, the time to detection, and certification costs from the planned demonstrations and to evaluate the applicability of DCGO to specific industries and regions quantitatively.

\section*{Acknowledgment}
This study was supported by the Japan CSTI BRIDGE Program with MIC, METI, and Digital Agency, and by JST PRESTO Grant Number JPMJPR2369.


\end{document}